# A Collective Anomaly Detection Method Over Bitcoin Network

Corresponding Author:

**Mohammad Javad Shayegan**

Department of Computer Engineering, University of Science and Culture

*University of Science and Culture, Bahar st, Shahid Qamushi st, Ashrafi Esfahani Bulvar*

Tehran, Iran

Academic email: shayegan@usc.ac.ir

ORCID**:** https://orcid.org/0000-0002-9635-2363

**Hamid Reza Sabor**

Department of Computer Engineering, University of Science and Culture

*University of Science and Culture, Bahar st, Shahid Qamushi st, Ashrafi Esfahani Bulvar*

Tehran, Iran

hr.saborroh@gmail.com

# A Collective Anomaly Detection Method Over Bitcoin Network

## Abstract

The popularity and amazing attractiveness of cryptocurrencies, and especially Bitcoin, absorb countless enthusiasts daily. Although Blockchain technology prevents fraudulent behavior, it cannot detect fraud on its own. There are always unimaginable ways to commit fraud, and the need to use anomaly detection methods to identify abnormal and fraudulent behaviors has become a necessity. The main purpose of this study is to present a new method for detecting anomalies in Bitcoin with more appropriate efficiency. For this purpose, in this study, the diagnosis of the collective anomaly was used, and instead of diagnosing the anomaly of individual addresses and wallets, the anomaly of users was examined, and the anomaly was more visible among users who had multiple wallets. In addition to using the collective anomaly detection method in this study, the Trimmed_Kmeans algorithm was used for clustering and the proposed method succeeded in identifying 14 users who had committed theft, fraud, and hack with 26 addresses in 9 cases. Compared to previous works, which detected a maximum of 7 addresses in 5 cases of fraud, the proposed method has performed well. Therefore, the proposed method, by presenting a new approach, in addition to reducing the processing power to extract features, succeeded in detecting abnormal users and also was able to find more transactions and addresses committed a scam.

## Keywords:

## 1-     Introduction

Blockchain was first proposed in 1991 to establish an encryption and information exchange system to address data security concerns [1]. Bitcoin as the first electronic cryptocurrency was emerged from the Blockchain features by Satoshi Nakamoto in 2008 [2] and attracted the attention of governments around the world to use Bitcoin. Attractiveness and the amazing popularity of Bitcoin as a cryptocurrency have made blockchain so popular. Blockchain has gained many enthusiasts in industry and academia and attracted the attention of many applications such as the Internet of Things [3]. Due to this popularity, many cybercriminals and even real-world criminals (because of the anonymity of users) became interested in using Blockchain and Bitcoin [4].

However, blockchain is not without drawbacks and limitations and is not completely immune to fraud, hack, attacks, and other malicious activities. Bitcoin users are always at risk of being hacked, and in addition to the enormous economic losses it causes to these users, it can also cause credit crises for commercial websites [5-8]. Due to this technology's novelty, the developed security mechanisms for some systems do not yet exist, and there have been several hack attacks on digital currencies [5]. Although Blockchain technology prevents fraudulent behavior, it cannot detect fraud on its own, so new innovative techniques and methods are needed to track attacks [9]. The amazing attractiveness of Bitcoin, on the one hand, and the rise of cybercrime activity, on the other, have made it imperative to use anomaly detection for identifying potential scams.

This study intends to use collective anomaly detection (instead of point anomaly detection) on all one user's wallets (instead of individual wallets) to remove features that have higher computational and operational capabilities. We believe this approach reduces data size and helps to better identify abnormalities that have been intentionally used with multiple user wallets.

## 2- Previous Works

Initially, blockchain was thought to be resistant to all kinds of attacks due to its cryptographic type and thanks to consensus algorithms, but security issues have prompted researchers to look for ways to detect anomalies in blockchain. Several studied tackled the anomaly detection issue in blockchain [4,8,10,6,11-18,7].

Table 1 shows a summary of the types of malicious attacks on Blockchains and the tactics and potential strategies that can be used to confront them. As shown in Table 1, anomaly detection methods can be used to detect the most malicious attacks. For example, using anomaly detection methods, bitcoin accounts of users who have used combinational services to engage in illegal activities or money laundering can be detected and tracked.

Table 1: summary of the types of malicious attacks on blockchain and potential strategies confront them (extracted from [19])

| **Malicious Attack** | **Definition** | **Defensive & Preventive Measures** |
|---|---|---|
| Double Spending | An individual makes more than one payment using one body of funds. | The complexity of the mining process |
| Record Hacking | Records in the ledger are modified or fraudulent transactions are inserted into the ledger. | Distributed consensus; Detection techniques |
| 51% Attack | A single miner node with more computational resources (51%) than the rest of the network nodes dominates the verification and approval of transactions. | Detection techniques; wide adoption of the blockchain technology |
| Identity Theft | The private key of an individual is stolen. | Identify and reputation blockchains |
| Illegal Activities | Parties transact illegal goods or commit money laundering. | Detection techniques; laws and regulations |
| System Hacking | The programming codes and systems that implement blockchain are compromised. | Robust systems and advanced intrusion detection methods |

According to Table 1, one of the diagnostic anomalies in blockchain is countering the Record Hacking attack and detecting theft, hacking, fraud, which has been considered by this study and the following works:

Zambre et al. [8] used six features and the K-Means algorithm to identify suspicious and rogue users and found a starting point for analyzing suspicious users. Pham et al. [10] used three main social networking methods (power degree & densification laws, K-Means clustering, and local outlier factor) to diagnose anomalies. They were able to discover one of the 30 known cases of theft. In a subsequent study, Pham et al. [6] used three unsupervised learning methods, including K-Means clustering, Mahalanobis distance, and unsupervised vector machine (SVM), and were able to identify a total of 3 of the 30 known cases.

Monamo et al. [4] emphasizing that anomaly detection plays an important role in data mining and considering that many remote locations have important information for further investigation, and in the Bitcoin network, diagnostics anomaly means detecting fraud, used the Trimmed-Kmeans method. They were able to successfully identify 5 of the addresses involved in 30 cases of theft, hack, fraud, or loss.

Monamo et al. used the kd-trees algorithm instead of the Trimmed-Kmeans algorithm in the next study [11] and were able to discover 7 of the target addresses, which were involved in 5 out of 30 cases of theft, hack, fraud, or loss. (In many cases of theft, hack, or fraud, thieves participated in multiple addresses and wallets to make it difficult to detect anomalies, leaving multiple addresses and wallets in each theft.)

In a study, Signorini et al. [12] suggested using Fork instead of eliminating it to diagnose abnormalities. Chawathe [13] further analyzes the method of Monamo et al. [4] and recommends this method to detect anomalies in the blockchain. In addition to the method and algorithm used in the previous records, the subject of feature selection is also very important, which is compared in Table 2.

According to the results obtained in previous records, several works have been conducted, and in most of these methods, improvements in anomaly detection have been achieved by changing or adding new features or changing the algorithm, but in all methods, only the anomaly detection of wallet addresses has been sought. And if the user has multiple wallets and the behavior of each of these addresses seems normal, the previous methods will be a bit inefficient. Since abnormal users mainly use multiple wallet addresses to normalize their behavior, it can be more efficient to choose a method that can examine the user's behavior instead of the wallet address. In order to solve this problem, in addition to using the best features and algorithms in the previous records, the method of collective anomaly detection has been used to pay more attention to the anomaly detection of users with several wallet addresses.

## 3- Research method

As mentioned in this research, the process of anomaly detection has been done with a collective anomaly approach. The details of the proposed method are described below:

### 3-1- Dataset and theft list

In this research, the dataset of the "ELTE Bitcoin Project" [20] has been used. Because in the previous records, the Bitcoin Blockchain dataset was used until April 7, 2013, the initial preprocessing was performed on the dataset. The list of addresses that have committed theft, fraud, hacking, or loss was then extracted.

Table 2: Comparison of feature selection in previous records

| Feature Selection | Zambre et al. [8] | Pham et al. [10] | Pham et al. [6], in the subsequent study | Monamo et al. [4] | Monamo et al. [11], in the subsequent study | Chawathe [13] |
|---|---|---|---|---|---|---|
| In-degree | | ✓ | ✓ | ✓ | ✓ | ✓ |
| Out-degree | | ✓ | ✓ | ✓ | ✓ | ✓ |
| Average amount incoming | | | | ✓ | ✓ | ✓ |
| Average amount outgoing | | | | ✓ | ✓ | ✓ |
| Average time interval between transactions | ✓ | ✓ | ✓ | | | |
| Average time interval between out transactions | ✓ | ✓ | ✓ | | | |
| Clustering coefficient | | | ✓ | ✓ | ✓ | ✓ |
| Average incoming speed | ✓ | ✓ | ✓ | | | |
| Average outgoing speed | ✓ | ✓ | ✓ | | | |
| In-acceleration | ✓ | | | | | |
| Out-acceleration | ✓ | | | | | |
| unique in-degree | | | ✓ | | | |
| unique out-degree | | | ✓ | | | |
| Balance | | | ✓ | | | |
| creation date | | | ✓ | | | |
| active duration | | | ✓ | | | |
| In-degree transaction | | ✓ | | | | |
| out-degree transaction | | ✓ | | | | |
| Total value of the transaction | | ✓ | | | | |
| Triangle | | | | ✓ | ✓ | ✓ |
| total amount sent | | | | ✓ | ✓ | ✓ |
| total amount received | | | | ✓ | ✓ | ✓ |
| standard deviation received | | | | ✓ | ✓ | ✓ |
| standard deviation sent | | | | ✓ | ✓ | ✓ |
| In-in | | | | ✓ | ✓ | ✓ |
| In-out | | | | ✓ | ✓ | ✓ |
| Out-in | | | | ✓ | ✓ | ✓ |
| Out-out | | | | ✓ | ✓ | ✓ |

### 3-2- Preprocessing

Since the best results in previous records are related to Monamo et al. [11], and they also used the 14 features listed in Table 1, in this study, investigations were performed on these features, and data preprocessing was conducted in the following three general steps:

- ✓ Data wiping: Records that have no input or output are removed. Consequently, the number of records is reduced from 13086527 to 10800406.
- ✓ Data aggregation and data size reduction:
  - To detect collective anomalies in this research, using the Contraction feature, all the addresses and wallets of a user are aggregated to extract the appropriate features according to this aggregation of data, and as a result, the number of records reached 5305678 records.

- Considering that there is a computational relationship between the two In-degree and Out-degree features, and according to the principle of data aggregation, two In-degree and Out-degree features can be eliminated compared to the method of Monamo et al. [11]
- Because in many thefts, hack, or fraud cases, the criminals work with multiple addresses and wallets to make it difficult to diagnose the anomaly, therefore in the previous works that used the method of point anomaly detection, two important features of Clustering coefficient and Triangle were used to extract better results by realizing the multiplicity of connections between these addresses and wallets. On the other hand, according to the new approach of this research in diagnosing collective anomalies of users (with multiple addresses and possible wallets) instead of identifying point anomalies of addresses and wallets, two Clustering coefficients and Triangle features which require which high computational and operational power can be removed for extraction.

✓ Data conversion: The min-max linear method was used to normalize the data.

### 3-3- Feature Extraction

The features of the proposed method are described in Table3.

Table 3: Describe the features of the proposed method in the Bitcoin network

| Feature | Definition |
|---|---|
| Average amount incoming | The average amount of bitcoins received to the address of the user's wallet |
| Average Amount outgoing | The average amount of bitcoins sent to the user wallet address |
| total amount sent | The total amount of bitcoins sent to the user's wallet address |
| total amount received | The total amount of bitcoins received to the address of the user's wallets |
| standard deviation received | The standard deviation of the number of bitcoins received to the address of the user's wallets |
| standard deviation sent | The standard deviation of the number of bitcoins sent to the user's wallet address |
| Average neighborhood ( In-in ) | The average neighborhood of inputs to inputs of all outputs |
| Average neighborhood (In-out ) | The average neighborhood of inputs to outputs of all outputs |
| Average neighborhood (Out-in) | The average neighborhood of outputs to inputs of all outputs |
| Average neighborhood (Out-out) | The average neighborhood of outputs to outputs of all outputs |

### 3-4- Trimmed K-means algorithm

One of the most important features of this algorithm is to place $\alpha^1$ percent of the outlier, which is very far from other clusters' centers in the 0 cluster. This feature is particularly important in the case of the considered problem and anomaly detection. Therefore, the Trimmed K-Means algorithm has been used for clustering, which is a more robust method than the basic K-Means method, and according to researches, will have better results [21].

### 3-5- Clustering

Due to a large number of records and data dimensions and also the reduction of clustering time, Monamo et al. [4] applied the clustering operation to one million records, but in the proposed method, due to data aggregation and size reduction, work was made on all records to extract more reliable results. In the proposed method, like the method of Monamo et al. [4], the value of $\alpha$ is equal to 0.01, and the value of $k^2$ is equal to 8. Furthermore, in the end, the results of all $9^3$ categories have been examined according to the list of pre-prepared thefts.

## 4- Findings

In this section, the results of the experiment are presented and discussed and then compared with previous works

### 4-1- Test results

As shown in Table 4, the proposed method uses the collective anomaly detection method for the first time compared to previous records. It succeeds in detecting anomalies of users who intend to show their behavior usually by having multiple wallet addresses, and the proposed method was successful in detecting 14 users with 26 addresses involved in 9 cases of theft, fraud, hack, or loss.

Table 4: Results of anomaly diagnosis with the proposed method

| Row | User number | Address number of wallet | Theft number | Theft name |
|---|---|---|---|---|
| 1 | 882066 | 882066 | 1 | Stone Man Loss |
| 2 | 3216635 | 3216635 | 3 | Stefan thomas loss |
| 3 | 913570 | 5034989 | 4 | Allinvain Theft |
| 4 | 149 | 5463950 | 6 | Mass MyBitcoin Thefts |
| 5 | 64 | 5125978 | 11 | October 2011 Mt. Gox Loss |
| 6 | 135 | 924292 | 14 | Linode Hacks |
| 7 | 135 | 1095327 | 14 | Linode Hacks |
| 8 | 135 | 2000790 | 14 | Linode Hacks |
| 9 | 135 | 2021669 | 14 | Linode Hacks |
| 10 | 135 | 2720178 | 14 | Linode Hacks |
| 11 | 135 | 4941747 | 14 | Linode Hacks |
| 12 | 135 | 5679585 | 14 | Linode Hacks |

---

[1] $\alpha$= Percentage of outliers from other clusters
[2] Number of clusters
[3] 9= Number of clusters (8) + (1) outliers from other clusters

| 13 | 731 | 827543 | 14 | Linode Hacks |
|---|---|---|---|---|
| 14 | 9538 | 3283795 | 14 | Linode Hacks |
| 15 | 9538 | 5295593 | 14 | Linode Hacks |
| 16 | 9538 | 5911894 | 14 | Linode Hacks |
| 17 | 1363830 | 2305801 | 14 | Linode Hacks |
| 18 | 1363830 | 3707950 | 14 | Linode Hacks |
| 19 | 1698477 | 1698477 | 17 | May 2012 Bitcoinica Hack |
| 20 | 1914 | 818018 | 23 | Bitfloor Theft |
| 21 | 1914 | 1740332 | 23 | Bitfloor Theft |
| 22 | 1914 | 4524766 | 23 | Bitfloor Theft |
| 23 | 1914 | 5517289 | 23 | Bitfloor Theft |
| 24 | 833694 | 833694 | 23 | Bitfloor Theft |
| 25 | 4212450 | 4212450 | 23 | Bitfloor Theft |
| 26 | 7083219 | 11225439 | 24 | Cdecker Theft |

As shown in Figure 1, the detected anomalous addresses are all in the 0 cluster, which is the same as the outliers and makes up exactly one percent of the total data.

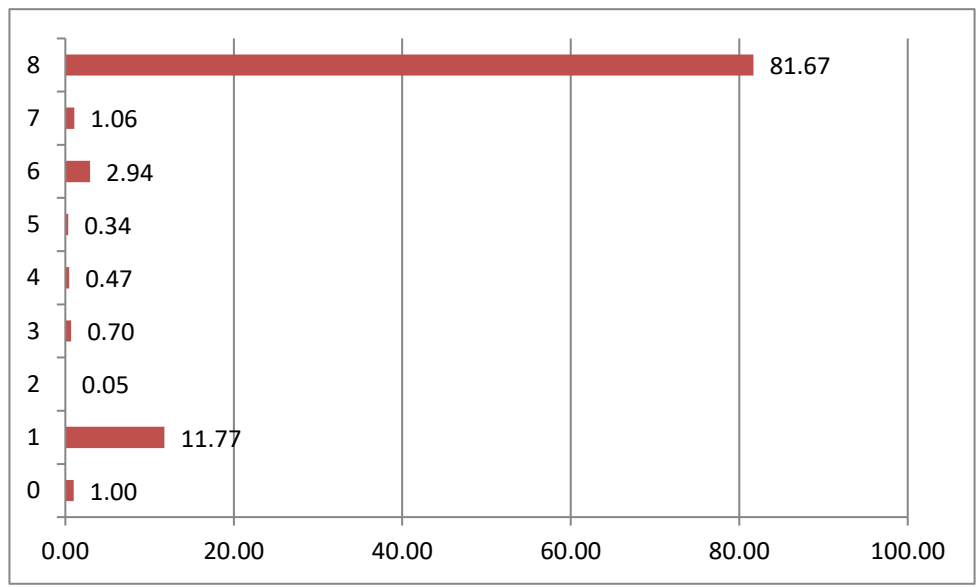

Figure 1: Dispersion rate of addresses after clustering in the proposed method

### 4-2- Comparison with previous works

The following section presents comparisons of the current study with the previous works. The comparions are based on features, employed algorithms, and the performance of the studies.

#### 4-2-1- Comparison of features

As shown in Table 5, the proposed method is placed in the middle of the table in terms of the number of extracted features. At the same time, considering that the proposed method does not use the clustering coefficient feature, the extraction of which has a high time complexity; therefore, the proposed method has acceptable performance in feature extraction in terms of computational and processing power.

Table 5: Comparison of the number of features, time, and computational and operational power

| Research name | Number of features | Approximate computational and operational time and power |
|---|---|---|
| Zambre [8] | 6 | Moderate |
| Pham [10] | 9 | Moderate |
| Pham et al. [6], in the subsequent study | 12 | High |
| Monamo [4] | 14 | High |
| Monamo [11], in the subsequent study | 14 | High |
| Chawathe [13] | 14 | High |
| The proposed method | 10 | Moderate |

### 4-2-2- Comparison of the used algorithms

As shown in Table 6, the proposed method was able to detect anomalies using only one algorithm and had a proper performance in selecting the algorithm.

Table 6: Comparison of used algorithms

| Study | The proposed algorithm |
|---|---|
| Zambre [8] | K-Means |
| Pham [10] | Power Degree & Densification Laws, K-Means Clustering, Local Outlier Factor |
| Pham [6] | K-Means, Mahalanobis, SVM |
| Monamo [4] | trimmed_kmeans |
| Monamo [11] | kd_tree |
| Chawathe [13] | trimmed_kmeans |
| The proposed method | trimmed_kmeans |

### 4-2-3- Comparison of success of the suggested approach

As shown in Table 7 and Figure 2, the proposed method identified 26 of the anomalous addresses that were present in the nine detected anomalies, and in this respect, performed better than the previous works.

Table 7: Types of thefts, hacks, scams, and losses detected by anomaly detection methods

| Theft Number | Name of theft, hack, fraud, loss | Pham [10] | Pham [6], in the subsequent study | Monamo [4] | Monamo [11], in the subsequent study | The proposed method |
|---|---|---|---|---|---|---|
| 1 | Stone Man Loss | | | | 1 | 1 |
| 3 | Stefan thomas loss | | | | | 1 |
| 4 | Allinvain Theft | | | | 1 | 1 |
| 5 | June 2011 Mt. Gox Incident | | | 1 | 1 | |
| 6 | Mass MyBitcoin Thefts | | | | | 1 |
| 11 | October 2011 Mt. Gox Loss | | | | | 1 |
| 14 | Linode Hacks | | | 3 | 3 | 13 |
| 17 | May 2012 Bitcoinica Hack | | | | | 1 |
| 23 | Bitfloor Theft | | | | | 6 |
| 24 | Cdecker Theft | | | | | 1 |
| 25 | 2012 50BTC Theft | | | 1 | 1 | |
| | Sum | 1 | 3 | 5 | 7 | 26 |

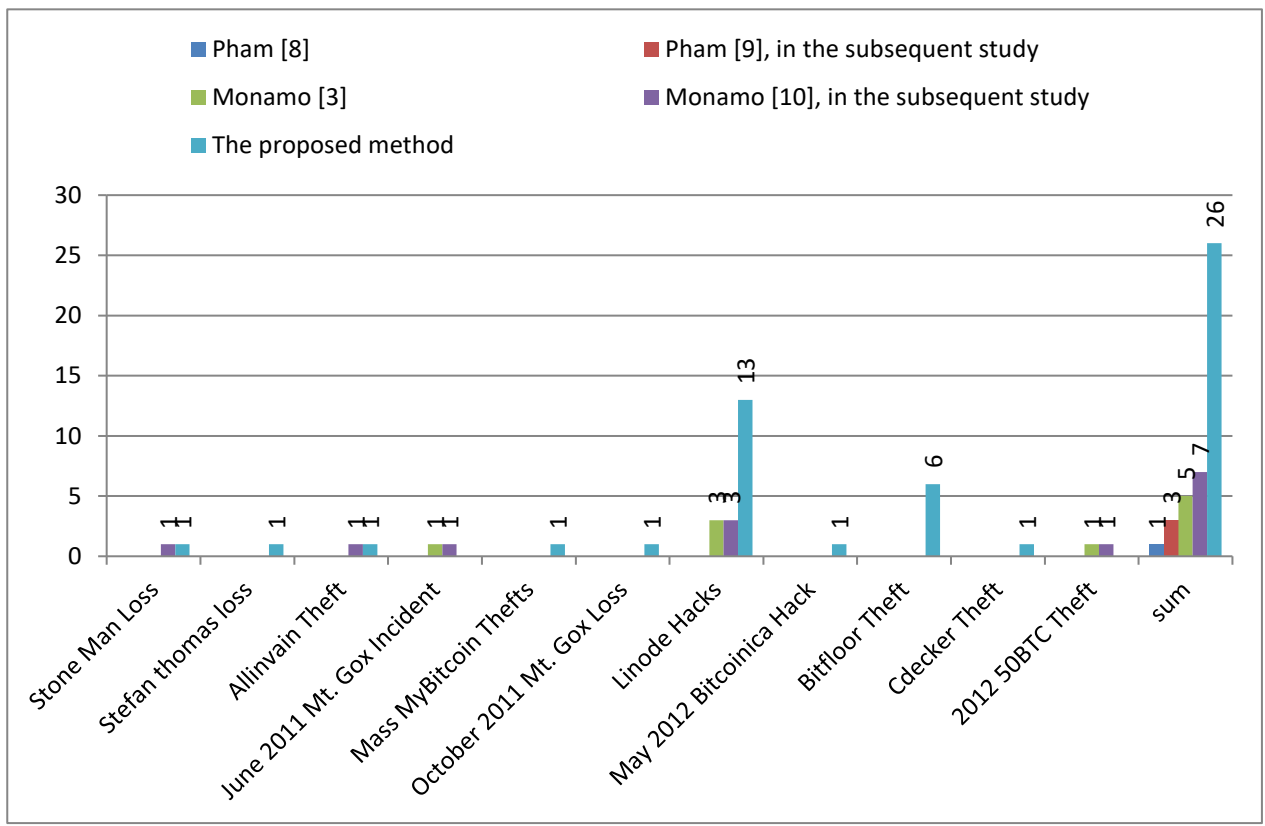

Figure 2: Thefts, hacks, scams, and losses detected by various Bitcoin anomaly detection methods

In terms of the number of features, the lowest number of features is related to Pham [10]. The proposed method is in the middle of the comparison table. Because the proposed method uses the diagnosis of collective anomalies, a reduction in the number of records has been created, and in general, it has been successful in reducing the dimensions (number of records and features).

In terms of time and operational and computational power, the proposed method performed better than previous records that managed to detect anomalies.

In terms of the number of algorithms used to detect anomalies and suspicious transactions, the proposed method using the Trimmed_Kmeans algorithm has performed finely.

The most important part of comparing the proposed method with others is the success in performance and results. In this regard, the proposed method has been able to achieve the best performance compared to other methods and was able to detect 14 users with 26 addresses (wallets) committed 9 cases of theft, fraud, hack, or loss, and compared to Monamo's latest method [11], which was able to find 7 addresses (wallets) that committed 5 thefts, scams, hacks, or losses has a much better performance.

## 5- Conclusion and suggestion

According to the results, it was found that people who intend to commit fraud and malicious activities in the Bitcoin network use several addresses and so-called digital wallets to normalize their activities as a normal user. In a way, these users' activity with multiple addresses makes them look almost like a normal user. To diagnose this type of anomaly, which is like an in-disguise anomaly, one must find a small deviation in these users' behavior. In the previous works, the anomaly detection was done by extracting new features that rely on the connection between a user's digital wallets. However, in the proposed method, using collective anomaly detection, the user's digital wallets are aggregated, and instead of detecting the anomaly of the wallet address, the anomaly of users who own one or more digital wallets was examined.

On the other hand, due to the significant expansion of this network, it becomes very difficult to extract features that depend on high power or computing time, and in practice, it seems very difficult to detect anomalies in this network with these methods. Therefore, to integrate and reduce the problem-solving dimensions of anomaly detection in Blockchain and Bitcoin networks, four features, two of which had high processing and computing power, were removed. The proposed method also uses the Trimmed_KMeans algorithm for clustering, which has a more robust method for solving anomaly detection problems than similar algorithms such as the KMeans algorithm. In the end, the proposed method was able to identify 14 users who had 26 known anomalous addresses. Thus, in comparison with the previous methods, in addition to reducing the dimensions of the problem from 10800406 records to 5305678 and also from 14 features to 10, the processing power and computational time of extracting each feature was also reduced. In addition, in the most important part of the evaluation and performance result, the number of detected thefts increased from 5 to 9 compared to the previous best methods, and the number of addresses of the perpetrators of theft was increased from 7 to 26. Also, in this method, for the first time, 14 users who committed these cases were identified.

As future works, it is suggested to do new work in two parts in general. In one step, features and algorithms should be selected that requires low computational and operational power to extract and execute. In another step, features and algorithms having the best diagnosis of the anomaly should be found.

**Declarations**

**Funding**: This research did not receive any specific grant from funding agencies in the public, commercial, or not-for-profit sectors**.**

**Conflicts of interest**: The authors declare that they have no conflict of interest.

**Availability of data and material**: The datasets generated during the current study are available from the authors on reasonable request.

**Code availability**: The code is available on demand

**Authors' contributions**: The authors declare that they contribute to all parts of this research and the extracted paper.